\documentclass[12pt]{article}
\usepackage{amssymb}
\usepackage{cite}
\usepackage{epsf}

\def\sgn{\mbox{sgn}}

\def\aaa{\alpha}
\def\bbb{\beta}
\def\ggg{\gamma}

\newcommand{\sfrac}[2]{{\textstyle\frac{#1}{#2}}}
\renewcommand{\thefootnote}{\fnsymbol{footnote}}

\begin{document}

\begin{titlepage}
\begin{flushright}
McGill 00-19\\
LBNL 46248\\
UCB-PTH-00/21\\
hep-th/0007029\\
\end{flushright}

\vskip.5cm
\begin{center}
{\huge{\bf Dynamical Instability of}}\\
\vskip.2cm
{\huge{\bf Brane Solutions with a }}\\
\vskip.3cm
{\huge{\bf Self-Tuning Cosmological Constant}}
\end{center}
\vskip1.5cm

\centerline{P.\ Bin\'etruy $^{a}$,  J.M.\ Cline $^{b}$ and  C. Grojean
$^{c,d}$}
\vskip 15pt               

\centerline{$^{a}$  LPT\footnote{Unit\'e mixte de recherche UMR n$^o$ 8627.},
Universit\'e Paris--XI, B\^atiment 211, F-91405 Orsay Cedex, France}
\vskip 3pt
\centerline{$^{b}$ Physics Department, McGill University,
Montr\'eal, Qu\'ebec, Canada H3A 2T8}
\vskip 3pt
\centerline{$^{c}$ Department of Physics,
University of California, Berkeley, CA 94720}
\vskip 3pt
\centerline{$^{d}$ Theoretical Physics Group,
 Lawrence Berkeley National Laboratory, Berkeley, CA 94720}
\vskip 3pt

\vglue .5truecm

\begin{abstract}
\vskip 3pt

\noindent
A five-dimensional solution to Einstein's equations coupled to a scalar
field has been proposed as a partial solution to the cosmological constant
problem: the effect of arbitrary vacuum energy (tension) of a 3-brane is
cancelled; however, the scalar field becomes singular at some finite
proper distance in the extra dimension. We show that in the original model
with a vanishing bulk
potential for the scalar, the solution is a saddle point which is unstable
to expansion or contraction of the brane world.  We construct exact
time-dependent solutions which generalize the static solution, and
demonstrate that they do not conserve energy on the brane; thus they do not
have an effective 4-D field theoretic description. When a bulk scalar field
potential is added, the boundary conditions on the brane cannot be
trivially satisfied, raising  hope that the self-tuning mechanism may
still give some insight into the cosmological constant problem in this
case. 

\end{abstract}

\end{titlepage}
\renewcommand{\thefootnote}{\arabic{footnote}}
\setcounter{footnote}{0}

\section{Introduction}

The idea that our world is a 3-brane embedded in extra spatial
dimensions has been widely discussed as a solution to the weak-scale
hierarchy problem \cite{AADD,RS}.  More recently, attention has been
focused on the possibilities for understanding the cosmological constant
problem within this setting \cite{Steinhardt,BMQ,VV,Schmidhuber1,ADKS,KSS1,KSS2,Youm,deAlwis1,CLP,LZ,
deAlwis2,Witten,BP,Schmidhuber2,FMRSW,Kakushadze1,TW,Kakushadze2,CPT,Krause}
(see \cite{brane-cc-old} for an earlier treatment).  A partial
solution was proposed in \cite{ADKS,KSS1} (ADKS--KSS), where the addition of a bulk
scalar, $\phi$, plays a crucial role (see also \cite{DWFGK,Gremm,Gubser}
for a discussion of the physics of a scalar field coupled to gravity).
The scalar is a free field in the
bulk, but has nontrivial couplings to fields living on the brane.  Ref.
\cite{ADKS} found static solutions of Einstein's equations with the
property that $\phi$ becomes singular at a finite distance in the extra
dimension, and the warp factor for the metric vanishes at the singularity.  
If one assumes that the extra dimension terminates at the singularity, or
that the warp factor remains integrably small beyond it, then gravity
appears to be four dimensional on large distance scales. Most importantly,
the scalar field is supposed to adjust itself to any arbitrary value of
the tension on the brane, which represents the four-dimensional
vacuum energy--or at least that part of it which comes from
nongravitational vacuum fluctuations.  The fact that the metric is static
means that the effective cosmological constant observed on the 3-brane
is zero, regardless of the size of the brane tension.  This could
constitute significant progress toward the solution of the cosmological
constant problem.

The self-tuning mechanism is incomplete in several ways.  In its original
form in \cite{ADKS,KSS1}, the orbifold solution requires a very particular
exponential coupling of $\phi$ to the matter fields on the brane,
$e^{\pm\kappa_5\phi}$, requiring just the right coefficient $\kappa_5$ in
the exponent, where $\kappa_5$ is related to the 5-D gravity scale $M_5$ by
$\kappa_5^2 = M_5^{-3}$.  As understood by ADKS, and explicitly realized in
refs.\ \cite{KSS2}, different choices of the coupling function $f(\phi)$
give de Sitter or anti-de Sitter branes in a $\mathbb{Z}_2$ bulk. 
Furthermore, the scalar potential in the bulk was assumed to vanish.  Ref.\
\cite{KSS1} extended the analysis to non-vanishing potentials, and
\cite{CEGH} gave the procedure for finding solutions with arbitrary
potentials in the bulk (see \cite{KOP} for an analytic solution
associated to a bulk cosmological constant and see
\cite{AJS,NOO,Zhu} for a discussion of
exponential potentials which can be associated with Neveu--Schwarz dilaton
tadpoles in non-supersymmetric string theories \cite{DM}).
Actually, as recently pointed out in Ref. \cite{FLLN2}, the
$\mathbb{Z}_2$ symmetric and 4D Poincar\'e invariant solution is unstable under
bulk quantum corrections: indeed with a conformal coupling
allowing flat solution with a vanishing bulk potential, the brane becomes
curved as soon as a bulk potential is turned on and the jump
equations relate the curvature of the brane to the value of the potential
on the brane by $R_{\mbox{\tiny 4D}}=\kappa_5^2 V(\phi_0)$. Supersymmetry
in the bulk may prevent from such instability.
Another
difficulty anticipated by \cite{ADKS,Youm,deAlwis1}, and explicitly shown
by \cite{FLLN1,FLLN2}, is that any procedure which regularizes
(``resolves'') the singularity in the solutions causes the reintroduction
of the fine-tuning which self-tuning is supposed to avoid, unless some
more explicit dynamical mechanism which automatically relaxes the effects
of the brane  tension can be demonstrated.
A further possible problem is the claim that when normal matter is added to
the brane tension, the brane remains static, in contradiction with
cosmology \cite{GNS} (however, see \cite{BHLLM} for a recent tentative
to recover usual cosmology).

In this letter we demonstrate a shortcoming which is more severe than the
foregoing ones; namely, starting from the very same Lagrangian which gives
the static self-tuned solutions, there also exist dynamical
solutions,
which either begin or end with a
singularity as time evolves.  In section 2 we will review the static
solution, and discuss the conformal symmetry which allows construction of
the dynamical solutions.  These constitute a family of solutions, of which
the static one is a special example.  In section 3 we emphasize that, even
starting arbitrarily close to the static solution within this family, the
brane world inevitably collapses to a singularity or else expands starting
from one, with a Hubble parameter of $H \sim \pm 1/(4t)$ as
$t\to\mp\infty$.  The interpretation is that the static solution
is a saddle point, unstable to small perturbations.  The solution on the
brane is shown to violate the positive energy condition, reflecting the
loss of energy from the brane into the bulk via the scalar field.  These
remarks apply in the case when the scalar bulk potential, $V(\phi)$,
vanishes.  In section 4 we show that for $V(\phi)\neq 0$, our
construction cannot be trivially applied to generate dynamical solutions. 
This gives further motivation for studying the stability of self-tuning
solutions with a nonvanishing scalar bulk potential. 


The solutions we constructed were independently found by Horowitz, Low and Zee
in Ref. \cite{HLZ} and were interpreted as describing a phase transition.
We will argue in the final section that they actually rather signal
an instability of the static solution.

\section{Dynamical Self-Tuning Solutions}

We will consider solutions arising from the action\footnote{Our conventions
correspond to a mostly positive Lorentzian
signature
$(-+\ldots +)$ and the definition of the curvature in terms of the metric is
such that a Euclidean sphere has  positive curvature. Bulk indices will be denoted
by capital Latin indices and brane indices by Greek indices:
$x^\mu$ are coordinates on the brane ($\tau$ or $t$ will be the time coordinate
and $x^i$ the spatial ones), and $y$ (or $z$, if the metric is
conformally flat) is the coordinate along the
fifth dimension such that
the brane is located at $y=0$ (or $z=0$).}
\begin{equation}
        \label{action}
S=\int d^{\,5}x\, \sqrt{|g_5|}
\left(\aaa R-\bbb
\nabla_M\phi \nabla^M\phi - \ggg V(\phi)\right)-
\int d^{\,4}x\,\sqrt{|g_4|}f(\phi_0)\,T,
\end{equation}       
where $g_5$ and $g_4$ are, respectively, the determinants the 5-D metric
$g_{MN}$ and the 4-D metric induced on the brane, $g_{\mu\nu}$, 
and $T$ is the bare tension.
The brane is supposed to couple to the bulk in a
conformal way defined by the function $f(\phi)$. The
physical tension is thus given by
\begin{equation}
        \label{tension}   
\tilde T =  f(\phi_0) T,
\end{equation}
where $\phi_0$ is the value of the scalar field on the brane.
There are many conventions for the
normalization of the terms in the bulk part of the action; to facilitate
comparison with other papers we will leave $\aaa,\bbb,\ggg$ unspecified.
We will be primarily concerned with the case of vanishing bulk potential,
$V(\phi)=0$, but we shall also consider nonzero $V(\phi)$ below.
Einstein's equations and the equation of motion for the scalar field
read
\begin{eqnarray}
&&
\alpha G_{MN} = \beta \nabla_M\phi \nabla_N\phi
-\sfrac{1}{2}\left( \beta (\nabla \phi)^2 + \gamma V(\phi) \right)g_{MN}
-\sfrac{1}{2}  f(\phi) T g_{\mu\nu} \delta^\mu_{M} \delta^\nu_{N}
\frac{\sqrt{|g_4|}}{\sqrt{|g_5|}}\, \delta (y) \ ;\\
&&
2 \beta \frac{1}{\sqrt{|g_5|}} \nabla_M \left(\sqrt{|g_5|}g^{MN}\nabla_N
\phi\right)
- \gamma \frac{d V}{d\phi} - \frac{d f}{d\phi} T
\frac{\sqrt{|g_4|}}{\sqrt{|g_5|}}\, \delta (y) = 0 \ ,
\end{eqnarray}
and, for a conformally flat metric with the form $ds^2 =
\Omega^2(z)(-d\tau^2+dx_i^2 + dz^2)$, 
the jump conditions for the derivatives of the fields at the brane are
\begin{equation}
	\label{eq:jumpRegular}
\left[ \Omega^{-2}\, \frac{d \Omega}{dz}\right]^{z=0^+}_{z=0^-}
= -\frac{T}{6\aaa} f(\phi_0)
\ \ \ \mbox{and} \ \ \
\left[ \Omega^{-1}\,  \frac{d \phi}{dz}\right]^{z=0^+}_{z=0^-}
=
\frac{T}{2\bbb} \frac{df}{d\phi} ( \phi_0).
\end{equation}

For a vanishing scalar bulk potential, the self-tuning solution of
\cite{ADKS,KSS1} with a $\mathbb{Z}_2$--symmetric bulk orbifold is given by
\begin{eqnarray}
	\label{eq:ADKS1}
&&
ds^2= \Omega^2(y) \, (-d\tau^2+dx_i^2) + dy^2
= \Omega^2(z)\,(-d\tau^2+dx_i^2+dz^2) \ ;\\
	\label{eq:ADKS2}
&&
\phi = \phi_0 \pm \sqrt{\sfrac{3\alpha}{4\beta}} \ln (1- |y|/y_c)
= \phi_0 \pm \sqrt{\sfrac{4\alpha}{3\beta}} \ln (1- |z|/z_c),
\end{eqnarray}
where $y$ is the proper distance coordinate, $z$ is the conformal
coordinate for the bulk and
\begin{equation}
\Omega(y) = \left( 1- |y|/y_c\right)^{1/4}, \; \; \Omega(z) =
\left( 1- |z|/z_c \right)^{1/3}.
\end{equation}
However, this solution satisfies the jump conditions (\ref{eq:jumpRegular})
only if the conformal
coupling is an exponential function with the particular form
\begin{equation}
 f(\phi) = \exp (\mp\sqrt{\sfrac{4\beta}{3\alpha}} \,\phi) .
\end{equation}
Then the integration constant, $y_c$ or $z_c$, is related to the brane tension by
\begin{equation}
y_c = \sfrac{3}{4} z_c = 3\aaa\,\tilde T^{-1} = 3\alpha
e^{\pm \sqrt{\sfrac{4\beta}{3\alpha}}\, \phi_0}T^{-1}
\end{equation}
and $\phi_0$ remains unconstrained---which is important for what follows.
A notable peculiarity of this conformal coupling is that it 
satisfies  the following equations everywhere in the bulk:
\begin{equation}
\label{eq:jumpY}
f(\phi(y)) = -\frac{12\alpha}{T}\, \sgn (y) \,\Omega^{-1}(y) \frac{d\Omega}{dy}
\ \ \mbox{and} \ \ \
\frac{df}{d\phi} (\phi(y)) = \frac{4\beta}{T}\, \sgn (y)\, \frac{d\phi}{dy}
\end{equation}
or, in conformal coordinates,
\begin{equation}
	\label{eq:jumpZ}
f(\phi(z)) = -\frac{12\alpha}{T}\, \sgn (z) \,\Omega^{-2}(z) \frac{d\Omega}{dz}
\ \ \mbox{and} \ \ \
\frac{df}{d\phi} (\phi(z)) = \frac{4\beta}{T}\, \sgn (z)\,\Omega^{-1}(z)
\frac{d\phi}{dz} .
\end{equation}
Hence, the relation between the field derivatives and $\phi$ required
by the jump conditions at the brane are satisfied not just there, but
everywhere in the bulk.

There is a simple procedure for transforming static bulk solutions
into dynamical ones. As emphasized in \cite{CGS4}, bulk symmetries
can be used to construct new solutions involving a singularity
interpreted as a brane. Starting from any regular static solution to
the bulk equations of motion, written for simplicity in conformal
coordinates,
\begin{equation}
	\label{sol-static}
\phi (z) \ \ \ \mbox{and} \ \ \ ds^2= \Omega^2 (z)\, \eta_{MN} dx^M \otimes dx^N ,
\end{equation}
we obtain a physically equivalent solution by applying a diffeomorphism
\begin{equation}
	\label{sol-diffeo}
{\tilde \phi} ({\tilde x}^\mu, {\tilde z}) = \phi (z({\tilde x}^\mu, {\tilde z}))
 \ \ \ \mbox{and} \ \ \
ds^2=  {\tilde g}_{MN} ({\tilde x}^\mu, {\tilde z}) \,
d{\tilde x}^M \otimes d{\tilde x}^N .
\end{equation}
However if we orbifold the new solution in the ${\tilde z}$ direction,
it becomes singular and describes a brane located at
${\tilde z}=0$. The orbifold projection and diffeomorphisms do not commute:
in general an orbifold solution constructed from (\ref{sol-diffeo})
is not equivalent under a change of coordinates
to an orbifold solution constructed from (\ref{sol-static}).
The difficulty consists in finding a diffeomorphism such that
the singularities introduced in the right hand side of the equations
of motion by the orbifold projection can be associated with the brane
components deduced from the action (\ref{action}).

In order to preserve the geometry of the brane embedded in the bulk,
we want to restrict ourselves to diffeomorphisms (\ref{sol-diffeo}) that
keep the metric diagonal,
\begin{equation}
 ds^2 = -{\tilde n}^2 ({\tilde x}^\mu, {\tilde z}) \, d{\tilde \tau}^2
+ {\tilde a}^2 ({\tilde x}^\mu, {\tilde z}) \, d{\tilde x}^2_i
+ {\tilde b}^2 ({\tilde x}^\mu, {\tilde z}) \, d{\tilde z}^2 .
\end{equation}
A particular subgroup is provided by the 5-D conformal transformations
under which the metric remains conformally flat
\begin{equation}
 ds^2 = {\tilde \Omega}^2 ({\tilde x}^\mu, {\tilde z})\,
\eta_{MN} \, d{\tilde x}^M \otimes d{\tilde x}^N .
\end{equation}
The infinitesimal conformal transformations are generated by the Killing
vectors
\begin{equation}
\xi^M = a^M + a^{[MN]}x_N + \lambda x^M + (x^P\! x_P \,\eta^{MN} - 2 x^M x^N)k_N
\end{equation}
where the parameters $a^M$, $a^{[MN]}$, $\lambda$,  $k_N$ correspond
respectively to translations, Lorentz transformations, dilations and special
conformal transformations.

In the present case, a combination of a boost in the $z$ direction and a
dilation provides a suitable diffeomorphism that will lead to a dynamical
solution also satisfying the boundary conditions on the brane. 
If $\phi(z)$ and $\Omega (z)$ is a regular solution in the bulk,
then it is simple to show that
\begin{equation}
{\tilde \phi} (z,\tau) = \phi (|z|+z_c h\tau)
\ \ \ \mbox{and} \ \ \
{\tilde \Omega}(z,\tau) = \frac{\Omega (|z|+z_c h\tau)}{\sqrt{1-z_c^2 h^2}}
\end{equation}
is a $\mathbb{Z}_2$-symmetric solution to the bulk equations of motion.
This can be checked from the explicit form of the bulk part of the
action (\ref{action}), which looks like
\begin{equation}
   \label{conf_action}
   S_{\rm bulk}=\int d^{\,5}x\,
   \left[4\aaa(\Omega(\partial\Omega)^2 + 2\Omega^2\partial^2\Omega) -\bbb
   \Omega^3(\partial\phi)^2 - \ggg \Omega^5 V(\phi)\right].
\end{equation}
Here the contractions of $\partial_M$ are performed with the Minkowski
space metric. This action has the additional symmetry that, for any
constant $\zeta$, leaves the equations of motion invariant:
\begin{equation}
\Omega \to \zeta\, \Omega;
\ \ \
\phi \to \phi;
\ \ \
V \to \zeta^{-2}\, V .
\end{equation}
In the case of a vanishing bulk potential, this symmetry implies
that
\begin{equation}
{\tilde \phi} (z,\tau) = \phi (|z|+z_c h\tau)
\ \ \ \mbox{and} \ \ \
{\tilde \Omega}(z,\tau) = \Omega (|z|+z_c h\tau)
\end{equation}
is a solution in the bulk.
Applying the procedure to the regular solution corresponding
to (\ref{eq:ADKS1})--(\ref{eq:ADKS2}), we obtain
\begin{eqnarray}
	\label{eq:dyn1}
&&
ds^2= \left( 1- |z|/z_c - h \tau \right)^{2/3}\,(-d\tau^2+dx_i^2+dz^2) \ ;\\
	\label{eq:dyn2}
&&
\phi = \phi_0 \pm \sqrt{\sfrac{4\alpha}{3\beta}} \ln (1- |z|/z_c - h\tau) .
\end{eqnarray}
The nontrivial step is to satisfy the jump
equations at the brane, which read
\begin{eqnarray}
        \label{jump1}
{\tilde \Omega}^{-2}\, \left.\frac{d{\tilde \Omega}}{dz}\right|_{z=0^+} =
-\frac{T}{12\aaa} \left.f({\tilde \phi}){\vphantom{\frac{d{\tilde
\phi}}{dz}}}\right|_{z=0^+};
\ \ \ \
{\tilde \Omega}^{-1}\,  \left.\frac{d{\tilde \phi}}{dz}\right|_{z=0^+} =
\frac{T}{4\bbb} \frac{df}{d\phi} ({\tilde \phi})
\left.{\vphantom{\frac{d{\tilde \phi}}{dz}}}\right|_{z=0^+} .
\end{eqnarray}
These equations are more difficult to satisfy when the solutions
are dynamical, because they must remain true for all conformal times
$\tau$. However, in the present case of a vanishing bulk potential, we
notice that, due to eqs.\ (\ref{eq:jumpY})--(\ref{eq:jumpZ}), the conformal
coupling
$f$ of
the scalar to the brane satisfies the relations (\ref{eq:jumpZ}) 
for any value of $z$, which ensures
that the dynamical solution we construct satisfies the jump equations
for any $\tau$, as can also be explicitly verified.
So from the same Lagrangian which gives the static self-tuned
solutions, there also exist dynamical solutions for which the induced
metric on the brane exhibits time dependence, as will be discussed
in the next section. The original
4-D Poincar\'e invariant solution corresponds to a very particular
value of the parameter $h$ that characterized our more general
family of solutions. This evades the no-go result of \cite{ADKS,KSS2}
which excluded the possibility of de Sitter or anti-de Sitter
branes in this case.

\section{Physical Interpretation}

The dynamical solution (\ref{eq:dyn1})--(\ref{eq:dyn2}) represents a singularity
which is either approaching or receding from the brane, depending on the
sign of the {continuous} parameter $h$.
{\it Assuming the extra dimension is simply truncated at the
singularity}, the strength of gravity is therefore time-dependent because
of the growth or collapse of the extra dimension.  The 4-D Planck mass
$M_p$ is related to the 5-D analogue $M_5$ by
\begin{equation}  
   \label{planck}
   M^2_p = 2 M_5^3 \int_{0}^{z_c(1-h\tau)}
   \Omega^3(z,\tau)\,  d z  = M_5^3 z_c (1 - h\tau).
\end{equation}

Moreover, an observer on the brane will see that his universe, although
spatially flat, does not
have 4-D Poincar\'e invariance, but it is growing or shrinking with a 
scale factor given by $\Omega(z=0,\tau)$.  In FRW time, $dt = \Omega
d\tau$, the scale factor and the corresponding Hubble parameter are
given by
\begin{equation}
	\label{FRW}
	a(t) = (1- h \tau)^{1/3} =  (1-{\sfrac43}h t)^{1/4};
\ \ \ \
	H = {\dot a\over a} = -{ h\over 3(1-{\sfrac43}ht)} .
\end{equation}
The universe either begins or ends in a singularity, depending on 
whether $h<0$ or $h>0$.  For the case $h=0$, the static solution of
ADKS is recovered.   Therefore one can interpret $h$ as the parameter 
determining how far away from the unstable saddle point solution one is,
in the space of all solutions.

The situation is qualitatively similar to a 4-D field theory analogy, in
which a cosmological ``constant'' $\Lambda$ is coupled to scalar fields
through the Lagrangian \footnote{We thank Nima Arkani-Hamed for making us
aware of this concept, through a related example.}
\begin{equation}
  \label{example}
  {\cal L} = a^3(t)\left[ {\sfrac12}(\dot\phi_1^2+\dot\phi_2^2)
	- \phi_1(\Lambda - \phi_2) \right] .
\end{equation}
This system also has a saddle point solution, at $\phi_2=\Lambda$ and
$\phi_1=0$, which could be construed as a self-tuning of the cosmological
constant to zero.  However, it is not a good solution to the cosmological
constant problem because it is unstable against small perturbations.  In 
figure 1(a) we show the time dependence of the Hubble parameter for both
the 5-D self-tuning solution and the 4-D toy model, in the case of a
collapsing universe, which is the generic outcome for the 4-D model.
Although $H(t)$ looks rather similar in the two cases, for the 4-D 
model $H$ starts positive and crosses zero, while for the 5-D
solution it is always negative.  The differences are more clearly seen
in figure 1(b), showing the scale factors $a(t)$ in FRW time.

\centerline{\epsfxsize=3.4in\epsfbox{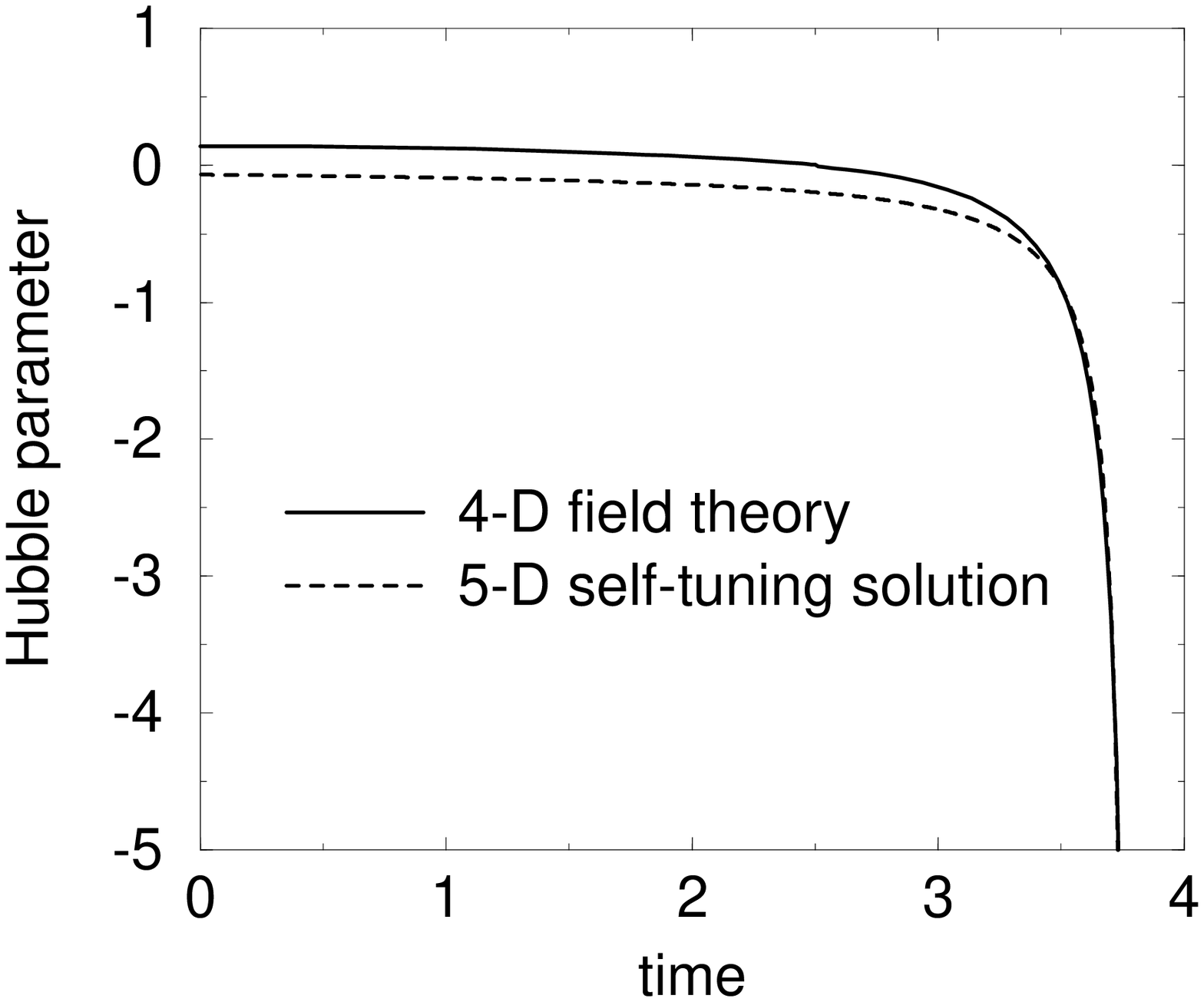}
\epsfxsize=3.4in\epsfbox{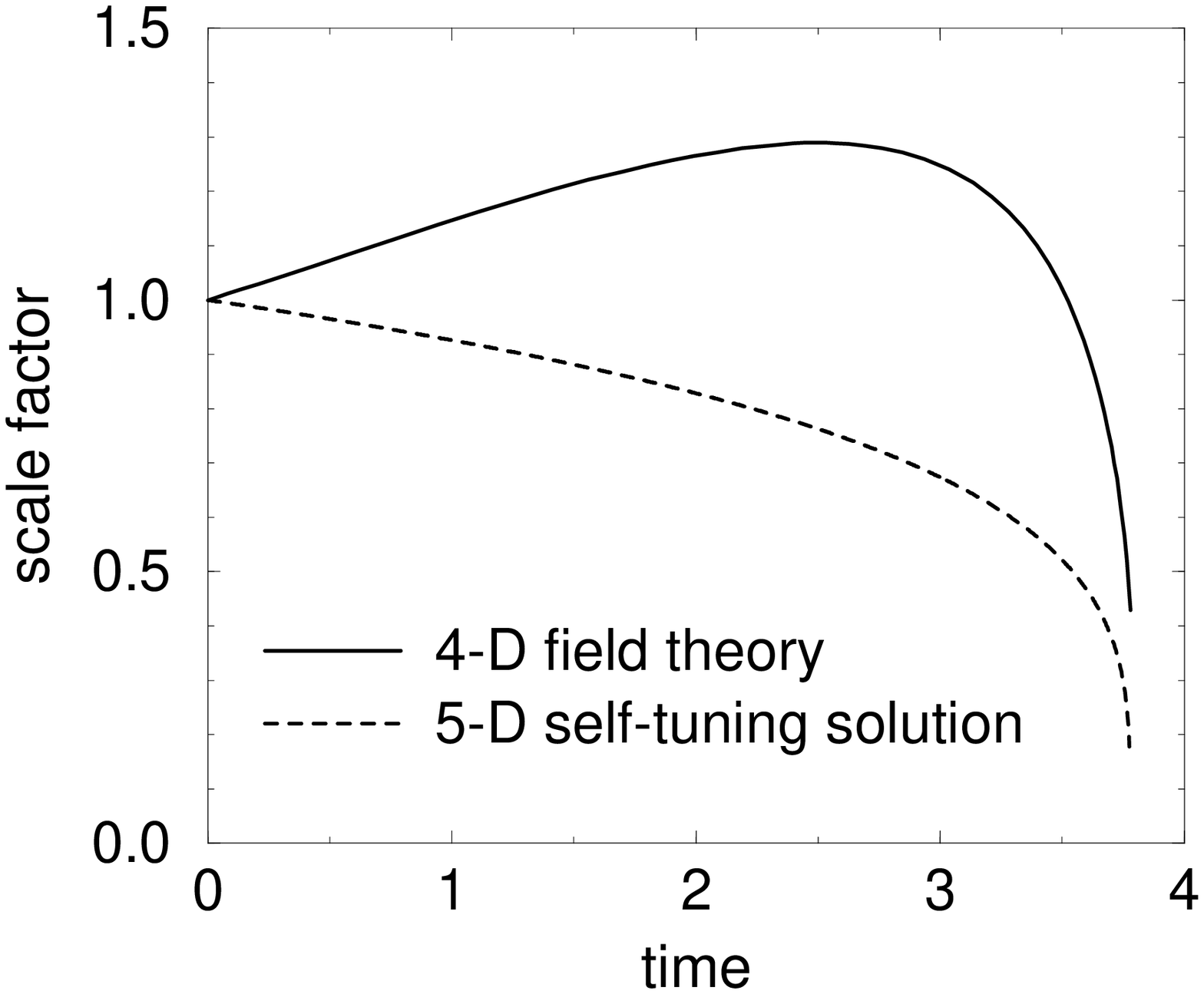}}
\noindent {\small Figure 1. (a) Comparison of the Hubble parameter as 
a function of time for the 5-D self-tuning solution and the 4-D toy
model (\ref{example}), for the case of a collapsing brane-world.
(b) Same, but showing the respective scale factors versus time. }
\vskip 0.5cm               

A further difference between the 5-D solution and any attempt to
describe it in a 4-D effective theory is that the 5-D solution
appears to violate energy conservation when viewed from the 3-brane.	
If we were in 4 dimensions, the scale factor dependence in
eq.\ (\ref{FRW}) would correspond to a
4-D stress energy tensor
\begin{equation}
	\label{stress}
	T^\mu_\nu = {\rho_0\over a^8(t)} 
	\left(\begin{array}{cccc} -1 & & & \\
				    & {\sfrac53} & & \\
				    & & {\sfrac53} & \\
				    & & & {\sfrac53}  
	\end{array}\right)
\end{equation}
where $\rho_0 = M_p^2 h^2/3$.  This has the equation of state
$p = (5/3)\rho$, which implies that there exists a null vector
$\xi_\mu$ which, when contracted with $T^\mu_\nu$, gives a spacelike
vector, contrary to the requirement of positivity of $T^\mu_\nu$.
This is due to the nonconservation of energy on the brane, which
can be explicitly demonstrated in the 5-D theory, by computing the
singular part of the divergence of the 5-D stress energy tensor\footnote{
In this equation $\rho$ and $p$ are the physical energy density and pressure
describing the matter living on the brane and conformally coupled to the
scalar field.}
\begin{equation}
          \label{conservation}
	\dot\rho + 3H(\rho+p) = {\sfrac14}(\rho-3p){f'\over f}
	\dot\phi
\end{equation}
In this context, $\rho-3p = 4\tilde T$, and in the dynamical solutions,
as long as $h$ is non-vanishing,
$\dot\phi$ is nonzero on the 3-brane, so the right hand side of
(\ref{conservation}) is nonzero.  More simply put, since the
physical tension ${\tilde T} = f(\phi) T$ is time-dependent,\footnote{
Curiously, if all the parameters of the Lagrangian (\ref{action}) as well as
the parameter $h$ are of the order of the 4-D Planck scale, with the
age of the Universe estimated to fifteen billions years, today the physical
tension would be of the order of $(5\, \mbox{TeV})^4$, close to
the electroweak scale.} energy is not conserved on the brane.

\section{Nonvanishing Bulk Potential}

An interesting question is whether our procedure can be generalized to the
case when the bulk potential is nonvanishing.  We do not have a definitive
answer, but we can argue that, if dynamical instabilities of the static
solution exist, they are much more difficult to find than when $V(\phi)=0$.
We are interested in orbifold solutions that can be constructed from a
regular bulk solution, $\Omega(z)$ and $\phi(z)$, which satisfies the jump
equations (\ref{eq:jumpRegular}) at $z=0$.  An important property of the
vanishing potential solution that facilitated finding dynamic solutions was
that the jump equations were actually satisfied for any $z$, not just at
$z=0$. This property cannot be maintained when the scalar potential is
turned on, while still having a self-tuning solution.  Indeed, it is easy
to verify that if the relations (\ref{eq:jumpRegular})  are satisfied for
any arbitrary values of $z$, the bulk equations of motion imply the
relation
\begin{equation}
V(\phi) = \frac{2}{\gamma}
\left( \frac{T^2}{32\beta} \left(\frac{df}{d\phi}\right)^2
-\frac{T^2}{24\alpha} f^2(\phi) \right)
\end{equation}
between the bulk scalar field and the conformal coupling.
This relation is incompatible with the self-tuning mechanism
unless the bulk potential vanishes; otherwise either $V(\phi)$ or $f(\phi)$
would have to depend on $T$, rather than $\phi_0$.
One way to overcome this difficulty
might be to consider diffeomorphisms $\tau,z\to\tilde\tau,\tilde z$ such
that ${\tilde z}=0$ implies $z=0$
independently of ${\tilde \tau}$. But it seems difficult to find
such a diffeomorphism that leaves the metric diagonal.
Another possibility would consist in relaxing the $\mathbb{Z}_2$ symmetry
in the bulk since a naive count of parameters \cite{CEGH} shows
that an integration constant remains unconstrained by the jump equations
and could be promoted to a time-dependent function. However the continuity
of the solution on the brane is no longer guaranteed at each time.
These observations may indicate that the self-tuned solutions
in the case of a nonvanishing bulk potential do not suffer
from the kind of instability we have found with the ADKS-KSS solution
whose pathology comes from the massless and unstabilized scalar field.

\section{Conclusion}

We have shown that the ADKS-KSS self-tuning solution is unstable
against eternal expansion or singular collapse of the brane-world.
Interestingly, the dynamics on the brane cannot be represented by a 4-D
field theory, since energy flows off the brane into the bulk and thus
appears not to be conserved on the brane.  We have suggested that these
problems may not occur in the presence of a potential energy in the
bulk for the scalar field.  If this hypothesis is correct, then further
indirect evidence would consist in perturbing the brane with matter or
radiation, and checking whether it behaves according to normal 4-D
cosmology.  We expect, in analogy to brane models without a stabilized
radion, that a nonstandard Friedmann equation on the brane, such as
$H\propto \rho$ \cite{BDL}, will be a diagnostic of instabilities in
the extra dimension, if they exist.  Work along these lines is
currently in progress \cite{BCG}.

In a closely related study, Horowitz, Low and Zee  recently presented
in \cite{HLZ} a general
class of plane wave solutions where the metric is parametrized as
$ds^2 = e^{2A(t,y)}(-dt^2+dy^2) + e^{2B(t,y)}dx_i^2$, and
$B(t,y) = (1/3)\ln(f(t-y) + g(t+y))$
for arbitrary functions $f$ and $g$.
The corresponding expressions for $A(t,y)$ and $\phi(t,y)$ are generally
more complicated, but it is possible to find solutions where
$A(t,y) = B(t,y)$ and $f$ and $g$ are both linear functions, which
are the same as our solutions.\footnote{When the warp factor is normalized to one on
the brane at $t=0$, the relation between the integration constants are
$z_c=y_0/(\xi-2)$, $h=\xi/y_0$ and $\phi_0=d+\epsilon \ln |y_0|$.}
Interestingly, \cite{HLZ} interpret these solutions as describing
a phase transition claiming that it is possible to start with the
static ADKS-KSS solution, at times $t<0$, and smoothly match it to the
dynamical solutions for $t>0$, provided there is a sudden
change $\Delta T$ in the brane tension at time $t=0$.  This would have
demonstrated a physical mechanism for triggering the instability: an
arbitrarily small change in the brane tension, such as would occur
during a first order phase transition.
However, it does not seem possible to have continuous time
derivatives of the fields when such a gluing of the two kinds of
solutions is attempted.\footnote{This can be seen in (V.18) of
\cite{HLZ}, where $y_*''(t)$ must have a Dirac delta function at $t=0$
if $y_*$ goes from being a constant to being linear in $t$ at $t=0$.}
Such a solution would require that the tension on the brane be proportional
to $\delta(t)$, rather than simply having a discontinuous change.
On the other hand, the property of having dynamical solutions evolving
to or from a Big Crunch or Bang with the {\it same} value of the brane tension
translates into an instability of the static solution with respect to
initial time derivatives since a small perturbation in
$\phi'_0$ drives the solution to a nonvanishing value
of $h$, and then unavoidably leads to a singularity in the time evolution.

\section*{Acknowledgements}

We thank Nima Arkani-Hamed, Csaba Cs\'aki, Joshua Erlich
and St\'ephane Lavignac
for stimulating discussions. P.B. thanks the Theory Group
at LBNL for its hospitality and its support.
C.G. is supported in part by the Director, Office of Science,
Office of High Energy and Nuclear Physics, of the US Department
of Energy under Contract DE-AC03-76SF00098,
and in part by the National Science Foundation under
grant PHY-95-14797.



\begin{thebibliography}{99}

\bibitem{AADD}
I.~Antoniadis,
{\it Phys.\ Lett.}\ {\bf B246} (1990) 377;
%
N.~Arkani-Hamed, S.~Dimopoulos and G.~Dvali,
{\it Phys.\ Lett.}\  {\bf B429} (1998) 263
{\tt [hep-ph/9803315]};
I.~Antoniadis, N.~Arkani-Hamed, S.~Dimopoulos and G.~Dvali,
{\it Phys.\ Lett.}\ {\bf B436} (1998) 257
{\tt [hep-ph/9804398]};
%
N.~Arkani-Hamed, S.~Dimopoulos and G.~Dvali,
{\it Phys.\ Rev.}\  {\bf D59} (1999) 086004
{\tt [hep-ph/9807344]};
%
N.~Arkani-Hamed, S.~Dimopoulos and J.~March-Russell,
{\tt [hep-th/9809124]};
%
{\it See also}
K.~R.~Dienes, E.~Dudas and T.~Gherghetta,
{\it Phys.\ Lett.}\  {\bf B436} (1998) 55
{\tt [hep-ph/9803466]}
%
and
{\it Nucl.\ Phys.}\  {\bf B537} (1999) 47
{\tt [hep-ph/9806292]}.


\bibitem{RS}
L.~Randall and R.~Sundrum,
{\it Phys.\ Rev.\ Lett.}\  {\bf 83} (1999) 3370
{\tt [hep-ph/9905221]}
and
{\it Phys.\ Rev.\ Lett.}\  {\bf 83} (1999) 4690
{\tt [hep-th/9906064]}.

\bibitem{Steinhardt}
P.~J.~Steinhardt,
{\it Phys.\ Lett.}\  {\bf B462} (1999) 41
{\tt [hep-th/9907080]}.

\bibitem{BMQ}
C.~P.~Burgess, R.~C.~Myers and F.~Quevedo,
{\tt hep-th/9911164}.

\bibitem{VV}
E.~Verlinde and H.~Verlinde,
{\it JHEP} {\bf 0005} (2000) 034
{\tt [hep-th/9912018]}.

\bibitem{Schmidhuber1}
C.~Schmidhuber,
{\tt hep-th/9912156}.

\bibitem{ADKS}
N.~Arkani-Hamed, S.~Dimopoulos, N.~Kaloper and R.~Sundrum,
{\tt hep-th/0001197}.
%

\bibitem{KSS1}
S.~Kachru, M.~Schulz and E.~Silverstein,
{\tt hep-th/0001206};

\bibitem{KSS2}
S.~Kachru, M.~Schulz and E.~Silverstein,
{\tt hep-th/0002121}.


\bibitem{Youm}
D.~Youm,
{\tt hep-th/0002147}.

\bibitem{deAlwis1}
%
S.~P.~de Alwis,
{\tt hep-th/0002174}.

\bibitem{Witten}
E.~Witten,
{\tt hep-ph/0002297}.


\bibitem{CLP}
J.~Chen, M.~A.~Luty and E.~Ponton,
{\tt hep-th/0003067}.

\bibitem{LZ}
I.~Low and A.~Zee,
{\tt hep-th/0004124}.

\bibitem{deAlwis2}
S.~P.~de Alwis, A.~T.~Flournoy and N.~Irges,
{\tt hep-th/0004125}.

\bibitem{BP}
R.~Bousso and J.~Polchinski,
{\it JHEP} {\bf 0006} (2000) 006
{\tt [hep-th/0004134]}.

\bibitem{Kakushadze1}
Z.~Kakushadze,
{\tt hep-th/0005217}.

\bibitem{Schmidhuber2}
C.~Schmidhuber,
{\tt hep-th/0005248}.

\bibitem{FMRSW}
J.~L.~Feng, J.~March-Russell, S.~Sethi and F.~Wilczek,
{\tt hep-th/0005276}.

\bibitem{Kakushadze2}
Z.~Kakushadze,
{\tt hep-th/0006059}.

\bibitem{TW}
S.~H.~Tye and I.~Wasserman,
{\tt hep-th/0006068}.


\bibitem{CPT}
A.~Chodos, E.~Poppitz and D.~Tsimpis,
{\tt hep-th/0006093}.

\bibitem{Krause}
A.~Krause,
{\tt hep-th/0006226}.


\bibitem{brane-cc-old}
V.A.~Rubakov and M.E.~Shaposhnikov,
{\it Phys. Lett.} {\bf 125B} (1983) 139.

\bibitem{DWFGK}
O.~DeWolfe, D.~Z.~Freedman, S.~S.~Gubser and A.~Karch,
{\tt hep-th/9909134}.

\bibitem{Gremm}
M.~Gremm,
{\tt hep-th/0002040}.

\bibitem{Gubser}
S.~S.~Gubser,
{\tt hep-th/0002160}.

\bibitem{CEGH}
C.~Cs\'aki, J.~Erlich, C.~Grojean and T.~Hollowood,
{\tt hep-th/0004133}, to be published in {\it Nucl. Phys.} {\bf B}.

\bibitem{KOP}
P.~Kanti, K.~A.~Olive and M.~Pospelov,
{\it Phys.\ Lett.}\  {\bf B481} (2000) 386
{\tt [hep-ph/0002229]}.

\bibitem{AJS}
N.~Alonso-Alberca, B.~Janssen and P.~J.~Silva,
{\tt hep-th/0005116}.

\bibitem{NOO}
S.~Nojiri, O.~Obregon and S.~D.~Odintsov,
{\tt hep-th/0005127}.

\bibitem{Zhu}
C.~Zhu,
{\tt hep-th/0005230}.

\bibitem{DM}
E.~Dudas and J.~Mourad,
{\tt hep-th/0004165}.

\bibitem{FLLN2}
S.~F\"orste, Z.~Lalak, S.~Lavignac and H.~P.~Nilles,
{\tt hep-th/0006139}.

\bibitem{FLLN1}
S.~F\"orste, Z.~Lalak, S.~Lavignac and H.~P.~Nilles,
{\it Phys.\ Lett.}\  {\bf B481}, 360 (2000)
{\tt [hep-th/0002164]}.


\bibitem{GNS}
B.~Grinstein, D.~R.~Nolte and W.~Skiba,
{\tt hep-th/0005001}.

\bibitem{BHLLM}
V.~Barger, T.~Han, T.~Li, J.~D.~Lykken and D.~Marfatia,
hep-ph/0006275.

\bibitem{HLZ}
G.~T.~Horowitz, I.~Low and A.~Zee,
{\tt hep-th/0004206}.

\bibitem{CGS4}
C.~Grojean,
{\it Phys.\ Lett.}\  {\bf B479} (2000) 273
{\tt [hep-th/0002130]}.

\bibitem{BDL}
P.\ Bin\'etruy, C. Deffayet and D. Langlois, 
{\it Nucl. Phys.} {\bf B565} (2000) 269
{\tt [hep-th/9905012]}.

\bibitem{BCG}
P.\ Bin\'etruy, J.M.\ Cline and C.\ Grojean, work in progress.

\end{thebibliography}
\end{document}